\newcommand{\be}{\begin{equation}}
\newcommand{\ee}{\end{equation}}
\newcommand{\bea}{\begin{eqnarray}}
\newcommand{\eea}{\end{eqnarray}}
\newcommand{\beann}{\begin{eqnarray*}}
\newcommand{\eeann}{\end{eqnarray*}}
\newcommand{\beasn}{\begin{sneqnarray}}
\newcommand{\eeasn}{\end{sneqnarray}}
\newcommand{\ba}{\begin{array}}
\newcommand{\ea}{\end{array}}
\newcommand{\nn}{\nonumber}
\newcommand{\Appendix}[1]%
    {\renewcommand{\thesection}{Appendix~\Alph{section}:}%
     \section{#1}}%

\catcode`@=11
\long\def\@makecaption#1#2{
   \vskip 10pt
   \setbox\@tempboxa\hbox{{\small\bf #1.} \ {\small #2}}
   \ifdim \wd\@tempboxa >\hsize       
   {\small\bf #1.} \ {\small #2}\par  
   \else                              
        \hbox to\hsize{\hfil\box\@tempboxa\hfil}
   \fi}
\catcode`@=12

\catcode`@=11
\def\secteqno{\@addtoreset{equation}{section}%
\def\theequation{\thesection.\arabic{equation}}}
\def\endsecteqno{\def\theequation{\@ifundefined{chapter}%
{\arabic{equation}}{\thechapter.\arabic{equation}}}}
\newcounter{subequation}
\def\thesubequation{\alph{subequation}}
\def\sneqnarray{\stepcounter{equation}\let\@currentlabel=\theequation
\setcounter{subequation}{1}
\def\@eqnnum{{\rm (\theequation\thesubequation)}}
\global\@eqcnt\z@\tabskip\@centering\let\\=\@eqncr\let\@@eqncr=\@@sneqncr
$$\halign to \displaywidth\bgroup\@eqnsel\hskip\@centering
 $\displaystyle\tabskip\z@{##}$&\global\@eqcnt\@ne
 \hskip 2\arraycolsep \hfil${##}$\hfil
 &\global\@eqcnt\tw@ \hskip 2\arraycolsep
$\displaystyle\tabskip\z@{##}$\hfil
  \tabskip\@centering&\llap{##}\tabskip\z@\cr}
\def\endsneqnarray{\@@sneqncr\egroup $$\global\@ignoretrue}
\def\@@sneqncr{\let\@tempa\relax
   \ifcase\@eqcnt \def\@tempa{& & &}\or \def\@tempa{& &}
   \else \def\@tempa{&}\fi
     \@tempa \if@eqnsw\@eqnnum\stepcounter{subequation}\fi
     \global\@eqnswtrue\global\@eqcnt\z@\cr}
\def\nobiblabels{\def\@lbibitem[##1]##2{\@bibitem{##2}}}
\catcode`@=12

\def\d{\delta}  

   \def\m{\mu} 
  \def\p{\pi}  
\def\s{\sigma} \def\t{\tau}

\def\pa{\partial}  

\documentclass[12pt]{article}
\usepackage[german,english]{babel}
\usepackage{amssymb}
\usepackage{amsmath}                                                         
\usepackage{epsfig,boxedminipage}
\usepackage{braket}

\secteqno
\renewcommand{\thesection}{\arabic{section}.}

\renewcommand{\theequation}{\thesection  \arabic{equation}}

\begin{document}


\title{{\bf Taking dibaryon fields seriously}} 
\author{{\Large {\sl Joan Soto}}  {\sl and} {\Large {\sl Jaume Tarr\'us}}\\
        \small{\it{Departament d'Estructura i Constituents de la Mat\`eria 
                   and Institut de Ci\`encies del Cosmos}}\\
        \small{\it{Universitat de Barcelona}}\\
        \small{\it{Diagonal, 647, E-08028 Barcelona, Catalonia, Spain.}}\\  \\
        {\it e-mails:} \small{tarrus@ecm.ub.es, joan.soto@ub.edu} }
\date{\today}

\maketitle

\thispagestyle{empty}


\begin{abstract}
We propose a low energy effective field theory of QCD at the scale of pion mass for the $N_B=2$ sector, $N_B$ being the baryon number, which contains two dibaryon fields in addition to the nucleons and pions. It has a well defined counting, is renormalizable and the nucleon-nucleon scattering amplitudes are manifestly unitary at leading order. We work out a lower energy 
effective theory for nucleons with energy much lower than 
the pion mass and three momentum comparable to it,
which also has a well defined counting and is renormalizable. The dibaryon fields must also be kept as explicit degrees of freedom in this theory. We calculate the scattering amplitudes 
at next-to-leading order 
for the $^1S_0$ and $^3S_1$ channels in this framework and obtain an excellent description of the phase shifts for center of mass energies in the $0-50 MeV$ range.

\end{abstract}

PACS: 14.20.Pt, 13.75.Cs, 21.30.Fe, 21.45.Bc, 03.65.Nk .

\vfill
\vbox{
\hfill{}\null\par
\hfill{UB-ECM-PF 07/33}\null\par}

\newpage


\section{Introduction}
\indent

Since the original suggestion by Weinberg \cite{Weinberg} that the nuclear forces could be understood within the framework of effective field theories (EFT) there has been an enormous development of the subject (see 
\cite{Bedaque:2002mn,Epelbaum:2005pn,Hammer:2006qj,Machleidt:2007ms}
for recent reviews). A key ingredient of the EFT formalism is that the cut-off dependence which is introduced in order to smooth out ultraviolet (UV) singularities can be absorbed by suitable counterterms, and hence any dependence on physical scales much higher than the ones of the problem at hand can be encoded in a few (unknown) constants. In order to achieve this in a systematic manner counting rules are also necessary.

Weinberg's suggestion consisted of two steps. The first one was calculating the nucleon-nucleon (NN) potentials order by order in Chiral Perturbation Theory
($\chi$PT) 
from the Heavy Baryon Chiral Lagrangian (HB$\chi$L) \cite{JM}. The second one introducing the potentials thus obtained in a Lippmann-Schwinger (LS) equation. There is no doubt that the first step can be carried out within an EFT framework: the renormalized NN potentials are known 
at leading order (LO), next-to-leading order (NLO), next-to-next-to-leading (NNLO) \cite{VK,Meissner} and next-to-next-to-next-to-leading \cite{Kaiser}, and isospin breaking terms have also been taken care of \cite{Walzl}. These potentials have been evaluated using static propagators for the nucleon fields. The use of non-relativistic propagators gives additional contributions starting at two loops, the leading order of which have been evaluated in \cite{Mondejar:2006yu}. The second step however is delicate. The potentials obtained in the first step are increasingly singular at 
short distances as we raise the order of $\chi$PT they are calculated. Hence the introduction of a regulator in the LS equation is compulsory. If a finite cut-off is accepted and allowed to move between a certain range, a very successful description of the phase shifts in different partial waves is achieved within this approach \cite{VK,Meissner,Entem:2003ft}. Nevertheless, even with the LO potential, 
it is not clear that the scattering amplitude thus obtained is cut-off independent.
An alternative to Weinberg's approach, which was free from renormalization problems, was proposed by Kaplan, Savage and Wise (KSW) \cite{KSW,Kaplan:1998we}. The key ingredient was to assume that contact interactions are enhanced with respect to the standard chiral counting. However, when the NN scattering amplitudes were worked out in this approach at NNLO a bad convergence of the series was observed, specially in the $^3S_1$ channel \cite{Stewart}. Since then a number of proposals
has been put forward \cite{Nogga:2005hy,Beane:2001bc}\cite{Entem:2007jg,PavonValderrama:2007nu,Valderrama:2005ku,PavonValderrama:2005uj,Valderrama:2005wv,PavonValderrama:2005gu,PavonValderrama:2004nb,PavonValderrama:2003np,Nieves:2003uu}\cite{Djukanovic:2006mc,Gegelia:2004pz,Gegelia:2001ev,Gegelia:1999gf,Gegelia:1998ee}\cite{Timoteo:2005ia,Frederico:1999ps}\cite{Yang:2007gk,Yang:2006ix,Yang:2004mq,Yang:2004ss,Yang:2004zg}\cite{Birse:2007sx,Birse:2005um}\cite{Oller:2003px}\cite{Eiras:2001hu}. Let us mention, for instance, ref. \cite{Beane:2001bc}
in which it is claimed that the renormalization program can actually be carried out in the Weinberg approach at LO if the potential is expanded about the chiral limit. However, the removal of the cut-off in this approach requires unconventional flows
and additional counterterms for higher partial waves \cite{Nogga:2005hy}.

Here we elaborate on the idea that the difficulties encountered so far in constructing a consistent and useful nucleon-nucleon effective field theory (NNEFT) may be a consequence of a misidentification of the low energy degrees of freedom. We will assume the NNEFT for energy and momentum scales much lower than $\Lambda_\chi$ contains two dibaryon fields as explicit degrees of freedom, with energy gaps (residual masses) of the order or smaller than the pion mass. If the dibaryon fields are naively integrated out, one gets the enhanced contact interactions of the KSW approach. We will argue that they must be kept as explicit degrees of freedom.

The relation between dibaryon fields and the KSW approach was noted early \cite{Kaplan:1996nv}. Dibaryon fields have also been used in EFT formulations of the three body problem (see \cite{Bedaque:1997qi,Bedaque:1999vb,Hammer:2007kq} and references there in). However, they have mostly been regarded as a convenient trick to carry out calculations (see, for instance, \cite{Beane:2000fi}). What it is new in our approach is the assumption that they {\it must} be included as explicit degrees in the NNEFT \footnote{In a model dependent framework, the inclusion of dibaryon fields as explicit degrees of freedom has already been advocated by some authors \cite{Kukulin:2005fv,Kukulin:2006wx,Pomerantsev:2005gw,Faessler:2005qq,Kukulin:2002sm})}. They cannot be integrated out if one wants to keep a natural counting. For a fundamental field theory their introduction should be irrelevant, since one can build the appropriate quantum numbers of the dibaryon out of the nucleon fields, and their inclusion does not affect the symmetries of the theory. For an effective theory, however, where calculations are necessarily organized in ratios of scales, it is extremely important to keep the appropriate degrees of freedom in the Lagrangian, even if they may appear redundant at first sight. We hope this will be illustrate in the paper with sufficient detail.

We will organize the paper as follows. In the next section we introduce the NNEFT with dibaryon fields, and discuss how the calculations must be organized. In section 3 we match it to a lower energy effective theory
for energies smaller than the pion mass and momenta comparable to it.
In section 4 we calculate the NN amplitudes at NLO. In section 5 we extract the low energy constants from data. In section 6 we critically examine the output of the previous section and propose new counting rules. In section 7 we discuss our results and section 8 is devoted to the conclusions.


\section{The nucleon-nucleon chiral effective theory with \\ dibaryon fields} 
\indent

We will consider an effective theory for $N_B$=2 sector of QCD for energies much smaller than $\Lambda_\chi$ about $2m_N$, $m_N$ being the nucleon mass. The usual degrees of freedom for such a theory, namely nucleons and pions, will be augmented by the inclusion of two dibaryon fields, an isovector ($D^a_s$) with quantum numbers $^1 S_0$ and an isoscalar ($\vec{D}_v$) with quantum numbers $^3 S_1$. Since $m_N \sim \Lambda_\chi$, a non-relativistic formulation of the nucleon fields is convenient \cite{JM}. Chiral symmetry, and its breaking due to the quark masses in QCD, constrains the possible interactions of the nucleons and dibaryon fields with the pions. The sector without dibaryon fields is the standard one \cite{Weinberg}, 

\be 
\begin{split}
\mathcal{L}_{\pi N}=& N^{\dag}\Bigl(iD_0-g_A(\vec{u}\cdot\frac{\vec{\s}}{2})+\frac{\vec{D}^2}{2m_N}\Bigr)N-\frac{C_S}{2}(N^{\dag}N)^2-\frac{C_T}{2}(N^{\dag}\vec{\s} N)^2+ \\
&+\frac{f_{\pi}^2}{8}\left\lbrace Tr(\pa_{\mu}U^{\dag}\pa^{\mu}U)+m^2_{\pi}Tr(U^{\dag}+U)\right\rbrace, \quad U=e^{2i\frac{\pi^a\t^a}{f_{\pi}}},
\end{split}
\ee
where $u^2=U$, $u_{\m}=i\left\lbrace u^{\dag},\pa_{\m}u\right\rbrace $, $D_{\m}=(\pa_{\m}+\frac{1}{2}[u^{\dag},\pa_{\m}u])$, $\pi^a$ is the pion field, $\t^a$ the isospin Pauli matrices, $g_A\sim 1.25$ is the axial vector coupling constant of the nucleon, and
$f_{\pi}\sim 132 MeV$ is the pion decay constant.
This is the leading order  Lagrangian for the pions ($\mathcal{O}(p^2)$) and the pion-nucleon interactions ($\mathcal{O}(p)$), augmented by the kinetic term of the nucleon (which is next-to-leading order for $E\sim p\sim m_\pi$, but becomes leading order
for $E\sim p^2/2m_N \ll m_\pi$).

The sector with dibaryon fields and no nucleons in the rest frame of the dibaryons reads
\be
\mathcal{L}_{D}=\mathcal{L}_{\mathcal{O}(p)}+\mathcal{L}_{\mathcal{O}(p^2)},
\ee
where $\mathcal{L}_{\mathcal{O}(p)}$ is the $\mathcal{O}(p)$ Lagrangian,
\be
\mathcal{L}_{\mathcal{O}(p)}={1\over 2} Tr\left[D_{s}^{\dag}\Bigl(-id_0+\delta_{m_s}'\Bigr)D_s\right]+
\vec{D}_v^{\dag}\Bigl(-i\pa_0+\delta_{m_v}'\Bigr)\vec{D}_v,
\label{dbLO}
\ee
where $D_s=D^a_s\t_a$ and $\delta_{m_i}'$, $i=s,v$ are the dibaryon residual masses, which must be much smaller than $\Lambda_\chi$, otherwise the dibaryon should have been integrated out as the remaining resonances have. The covariant derivative for the scalar (isovector) dibaryon field is defined as
$d_0D_s=\pa_0D_s+\frac{1}{2}[[u,\pa_0u],D_s]$.
$\mathcal{L}_{\mathcal{O}(p^2)}$ is the $\mathcal{O}(p^2)$ Lagrangian,
\be
\begin{split}
\mathcal{L}_{\mathcal{O}(p^2)}=
&s_1Tr[D_s(u\mathcal{M}^{\dag}u+u^{\dag}\mathcal{M}u^{\dag})D^{\dag}_s]+s_2Tr[D^{\dag}_s(u\mathcal{M}^{\dag}u+u^{\dag}\mathcal{M}u^{\dag})D_s]+\\
&+s_3Tr[D^{\dag}_sD_su_0u_0]+s_4Tr[D_sD^{\dag}_su_0u_0]+s_5Tr[D^{\dag}_sD_su_iu_i]+\\
&+s_6Tr[D_sD^{\dag}_su_iu_i]+s_7Tr[D_s^{\dag}u_0D_su_0]+s_8Tr[D_s^{\dag}u_iD_su_i]+\\
&+v_1\vec{D}^{\dag}_v\cdot\vec{D}_vTr[u^{\dag}\mathcal{M}u^{\dag}+u\mathcal{M}^{\dag}u]+v_2\vec{D}^{\dag}_v\cdot\vec{D}_vTr[u_0u_0]+\\
&+v_3\vec{D}^{\dag}_v\cdot\vec{D}_vTr[u_iu_i]+v_4(D^{i\dag}D^{j}+D^iD^{j\dag})Tr[u_iu_j],
\end{split}
\label{ordrep2}
\ee
$\mathcal{M}$ is the quark mass matrix, which we will take in the isopin limit, namely the average of the up and down quark masses $m_q$ times the identity matrix. $s_i$, $i=1,..,8$ and $v_j$, $j=1,..,4$ are low energy constants (LEC).

The dibaryon-nucleon interactions will only be needed at leading order
\be
\begin{split}
\mathcal{L}_{DN}=&\frac{A_s}{\sqrt{2}}(N^{\dag}\s^2\t^a\t^2N^*)D_{s,a}+\frac{A_s}{\sqrt{2}}(N^{\top}\s^2\t^2\t^aN)D^{\dag}_{s,a}+\\
&+\frac{A_v}{\sqrt{2}}(N^{\dag}\t^2\vec{\s}\s^2N^*)\cdot\vec{D}_v+\frac{A_v}{\sqrt{2}}(N^{\top}\t^2\s^2\vec{\s} N)\cdot\vec{D}_v^{\dag},
 \end{split}
\label{dn}
\ee
$A_i\sim \Lambda_\chi^{-1/2}$, $i=s,v$.

The nucleon-nucleon scattering amplitude will be dominated by the dibaryon field. At tree level it gives a contribution $\sim 1/m_\pi \Lambda_\chi$ (for energies $\sim m_\pi$) which is parametrically larger than the contribution arising from the four nucleon interactions $\sim 1/\Lambda_\chi^2$. The dibaryon field propagator gets an important contribution to the self-energy due to the interaction with the nucleons 
(Fig.1b),
which is always parametrically larger than the energy $E$. As a consequence the LO expression for the dibaryon field propagator becomes (in dimensional regularization (DR) and minimal subtraction (MS) scheme) ,
\be
\frac{i}{\d_{m_i}'+i\frac{A_i^2m_Np}{\pi}},
\label{dbself}
\ee
($p=\sqrt{m_N E}$) rather than the tree level expression $i/(-E+\delta_{m_i}'-i\eta)$. Note that (\ref{dbself}), unlike the tree level expression, has the correct positivity properties irrespectively of the sign chosen for the time derivative in (\ref{dbLO}). The unconventional signs for the time derivatives in (\ref{dbLO}) are chosen in this way in order to correctly reproduce the sign of the effective ranges later on. They do not imply any violation of unitarity because the correct leading order expression for the propagator is (\ref{dbself}) and not the tree level one. From (\ref{dbself}) it follows that the leading contribution to the NN scattering amplitude for energies $\sim m_\pi$ is parametrically $\sim 1/m_\pi^{1/2} \Lambda_\chi^{3/2}$, namely slightly suppressed with respect to the tree level estimate for it, but still more important than the tree level contribution from the four nucleon interactions ($\sim 1/\Lambda_\chi^2$).  The fact that a loop contribution always dominates over a tree level one is kind of bizarre in an EFT framework. It probably indicates that the first reliable approximation to the true amplitude is the NLO one, namely the first order in which the tree level energy dependence is not neglected. It also suggests that NLO contributions (and probably beyond) should be resummed in some sort of self-energy. We will see later on that unitarity provides the key ingredient to carry out these resummations. 

Equation (\ref{dbself}) implies that the dibaryon field should not be integrated out unless $p\ll \d_{m_i}'$, instead of $E\ll \d_{m_i}'$ as the tree level expression suggests. If $\d_{m_i}' \ll m_\pi$, it should also be kept as an explicit degree of freedom in the so called pionless EFT, like in Refs. \cite{Ando:2004mm,Ando:2005dk,Ando:2005cz}. 

Except for the above mentioned contributions to the self-energy of the dibaryon fields, which become LO, the calculation can be organized perturbatively in powers of $1/\Lambda_\chi$. Hence one expects that any UV divergence arising in higher order calculations will be absorbed in a low energy constant of a higher dimensional operator built out of nucleon, dibaryon and pion fields (note that the linear divergence in the self-energy of the dibaryon fields due to the diagram in Fig.1b can be absorbed in $\delta_{m_i}$).
The renormalized result may be recast in a manifestly unitary form \cite{Oller:2003px,Oller:2000ma}.

We shall restrict ourselves in the following to energies $E\sim m_\pi^2/\Lambda_\chi \ll m_\pi$, which implies nucleon three momenta $\sim m_\pi$. We shall follow the strategy of \cite{Eiras:2001hu}, which was inspired in the formalism of \cite{Mont}, and shall build a lower energy EFT with no explicit pion fields: the effects due to the pions will be encoded in the potentials (and redefinitions of the LECs). We will present the calculation of the NN scattering amplitudes at NLO. For these energies the LO contribution is $\sim 1/m_\pi \Lambda_\chi$, which means that we are aiming at including all contributions up to $1/\Lambda_\chi^2$.  
  
\section{The lower energy effective theory}

For energies $E\sim m_\pi^2/\Lambda_\chi \ll m_\pi$, the pion fields can be integrated out. This integration produces nucleon-nucleon potentials and redefinitions of low energy constants. Since we are aiming at a NLO calculation we must keep corrections $O(m_\pi/\Lambda_\chi)$ and neglect higher order ones.

In the one nucleon sector pion loops produce corrections which are $O(m_\pi^2/\Lambda_\chi^2)$ and hence can be neglected. The same holds true for 
the dibaryon sectors. 
However, both single nucleon and dibaryon sectors get a contribution $O(m_\pi/\Lambda_\chi)$ from counterterms proportional to the quark masses which redefine the nucleon mass and the dibaryon residual masses. 

In the two nucleon sector, the one pion exchange is the only relevant contribution at this order, which produces the well known one pion exchange (OPE) potential.

Pion loops in the dibaryon-nucleon vertices also produce $O(m_\pi^2/\Lambda_\chi^2)$ corrections, except for those which reduce to the OPE potential correction to the dibaryon-nucleon vertex, which are included in the effective theory and must not be considered in the matching.

Hence the Lagrangian of the lower energy effective theory at the NLO order reads as follows.
The sector without dibaryon fields reduces to the LO one in the Weinberg approach \cite{Weinberg}, 
\be
\begin{split}
\mathcal{L}_{\pi N}=& N^{\dag}\Bigl(i\pa_0+\frac{\vec{\pa}^2}{2m_N}\Bigr)N-\frac{C_S}{2}(N^{\dag}N)^2-\frac{C_T}{2}(N^{\dag}\vec{\s} N)^2+\\
&+\frac{1}{2}N^{\dag}\s^i\vec{\t}N(x_1)V_{ij}(x_1-x_2)N^{\dag}\s^j\vec{\t}N(x_2), 
\end{split}
\ee
where $V_{ij}$ is the OPE potential,

\be
V_{ij}(x_1-x_2)=-\frac{g^2_A}{2f^2_{\pi}}\int\frac{d^3q}{(2\pi)^3}\frac{q_iq_j}{\vec{q}^2+m^2_{\pi}}e^{-i\vec{q}\cdot(\vec{x}_1-\vec{x}_2)}.
\ee
The sector with dibaryon fields and no nucleons in the rest frame of the dibaryons reads
\be
\mathcal{L}'_{D}=D_{s,a}^{\dag}\Bigl(-i\pa_0+\delta_{m_s}\Bigr)D^a_s+
\vec{D}_v^{\dag}\Bigl(-i\pa_0+\delta_{m_v}\Bigr)\vec{D}_v,
\ee
$\delta_{m_i}$, $i=s,v$ is the (redefined) dibaryon residual mass, 
\bea
\delta_{m_s}&=& \delta_{m_s}'+4m_q(s_1+s_2)-2\delta m_N\nn\\
\delta_{m_v}&=& \delta_{m_v}'+4m_q v_1-2\delta m_N,
\label{shift}
\eea
$\delta m_N$ is also proportional to $m_q$ and stands for the leading $m_q$ corrections to nucleon mass (see, for instance, \cite{Bernard:1995dp}), which can be reshuffle into $\delta_{m_i}$ by local field redefinitions. $s_1$, $s_2$ and $v_1$ are the LEC introduced in equation (\ref{ordrep2}). Note that if $\d_{m_i}'\ll m_\pi$ then the quark mass dependence of $\d_{m_i}$ is a leading order effect.
The dibaryon-nucleon interactions remain the same as in (\ref{dn}).

The calculations in this EFT can be organized in ratios $E/p$ and $p/\Lambda_\chi$ (recall $m_\pi\sim p$). The UV divergences arising at higher orders will be absorbed by local terms build out of nucleon and dibaryon fields.

\section{The scattering amplitudes at NLO}

\begin{figure}
\centerline{
\begin{tabular}{cc}
\resizebox{5cm}{2.75cm}{\includegraphics{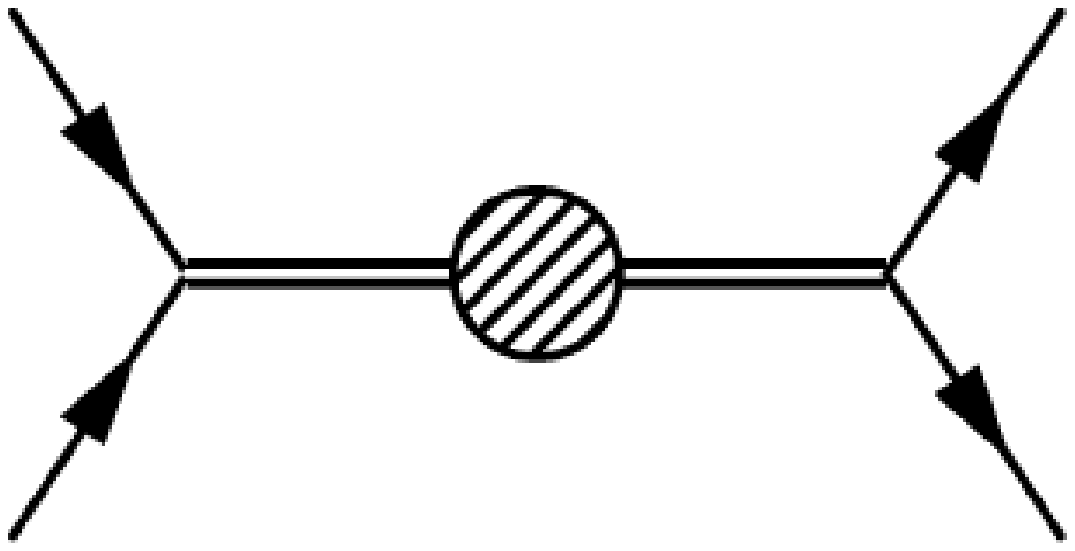}}& \quad
\resizebox{4cm}{2.75cm}{\includegraphics{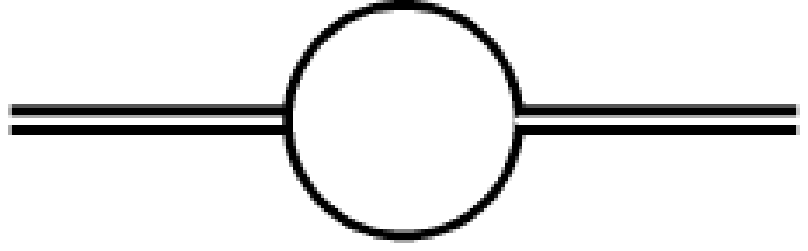}} \\
(a) & (b)
\end{tabular}
}
\caption{{\footnotesize a) LO diagram. The shaded circle stands for the resummation of dibaryon selfenergy. b) Dibaryon selfenergy diagram.}}
\label{fig1}
\end{figure}  

The scattering amplitudes at LO are given by the diagrams in Fig.1. The corresponding amplitudes are, 
\be
\mathcal{A}^{i}_{-1}=-\frac{1}{1+i\frac{A^2_im_Np}{\pi \d_{m_i}}}\frac{4A^2_i}{\d_{m_i}} \qquad i=s,v ,
\ee
Subscripts are $i=s,v$, and superscripts (which appear below) are $i=$ $^1S_0$, $^3S_1$ with the equivalence $s=$ $^1S_0$, $v=$ $^3S_1$. We use DR and the MS scheme throughout. 
These amplitudes formally coincide with the leading order of the effective range expansion (ERE) with the scattering length given by $a_i=\frac{A_i^2m_N}{\pi\d_{m_i}}$, and hence they also coincide with the LO amplitudes in the KSW approach identifying $\frac{\d_{m_i}}{4A_i^2}$ with $\frac{1}{C^i_0}+\frac{m_N\mu}{4\pi}$.
The bubble resummation chain in the latter plays the role of the dibaryon field. Notice, however, that the use of the dibaryon field makes unnecessary the use of the PDS scheme in order to keep a consistent counting for $p\sim m_\pi$. Furthermore, one must keep in mind that $\d_{m_i}$ may contain a leading order dependence on the quark masses if $\d_{m_i}\ll m_\pi$, whereas the LO scattering lengths in the KSW approach do not depend on the quark masses.

\begin{figure}
\centerline{
\begin{tabular}{ccc}
\resizebox{3.5cm}{2.75cm}{\includegraphics{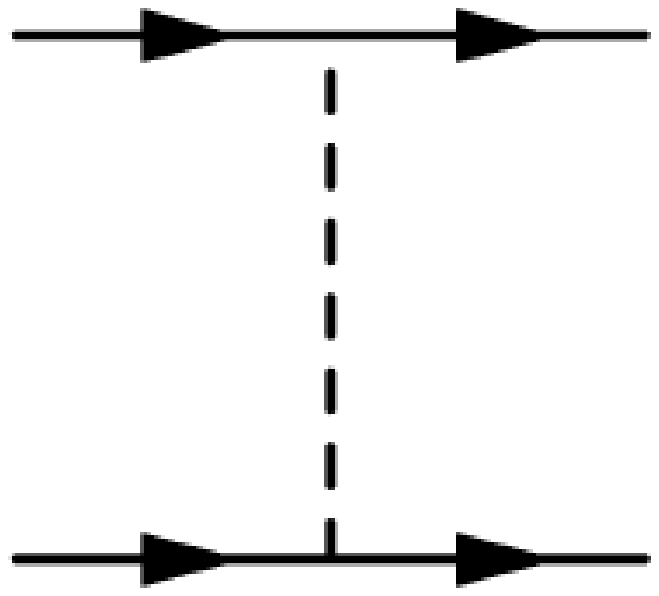}}& \quad
\resizebox{5cm}{2.75cm}{\includegraphics{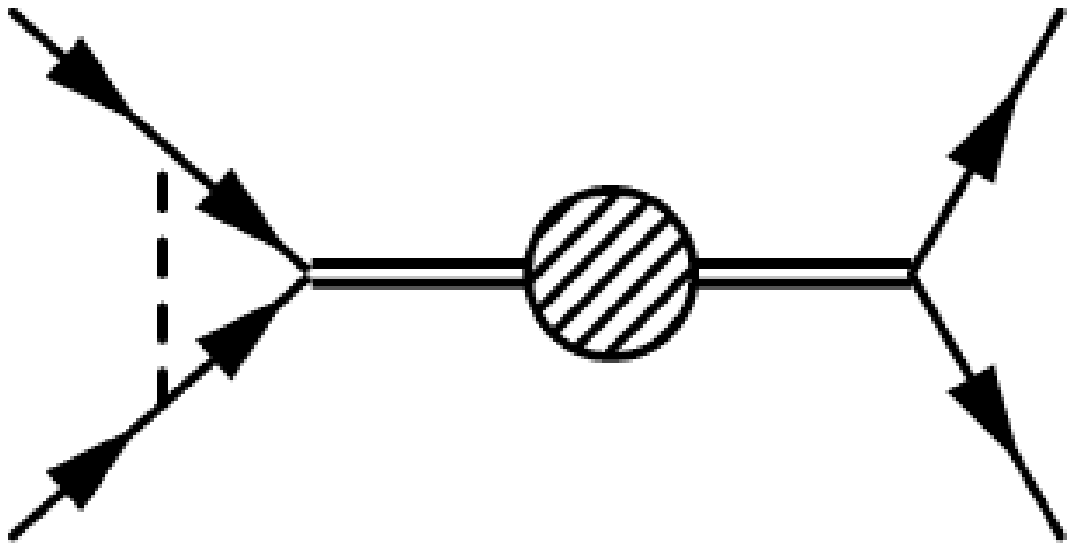}}& \quad
\resizebox{5cm}{2.75cm}{\includegraphics{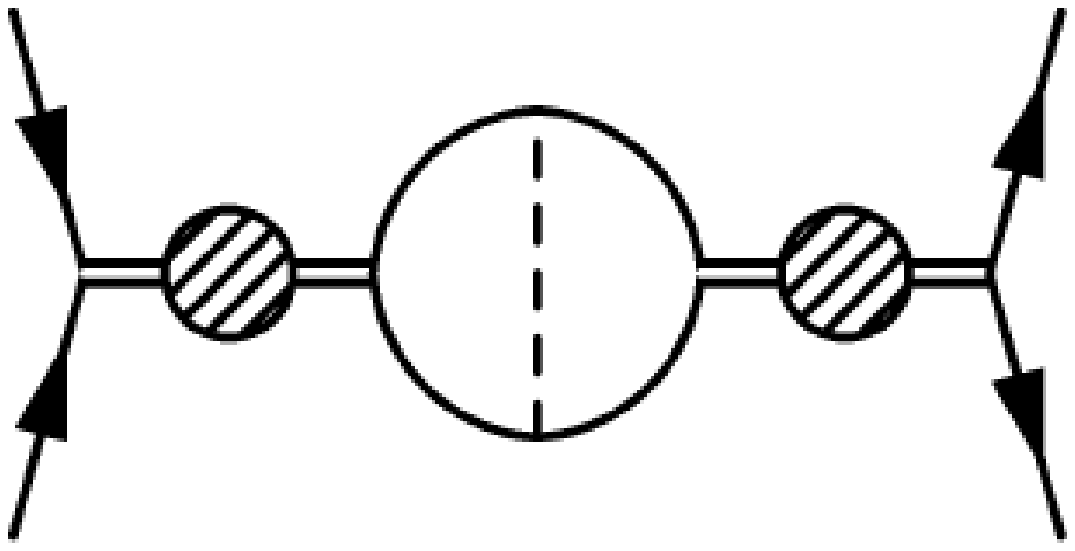}} \\
&\quad(a)&\quad \\ 
\resizebox{3.5cm}{2.75cm}{\includegraphics{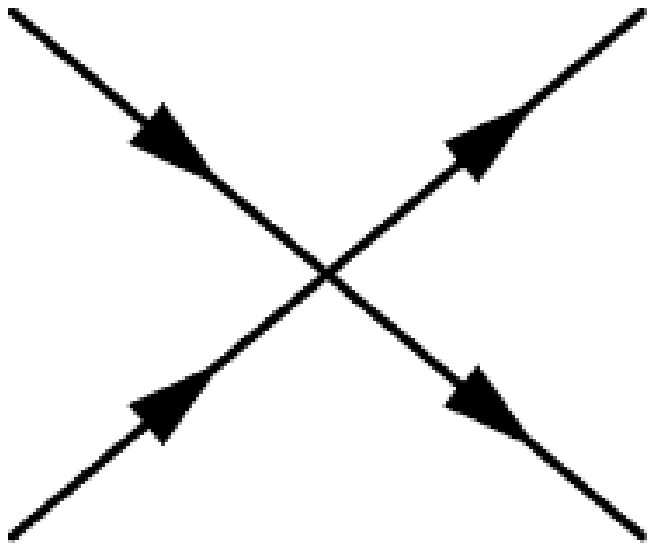}}& \quad 
\resizebox{5cm}{2.75cm}{\includegraphics{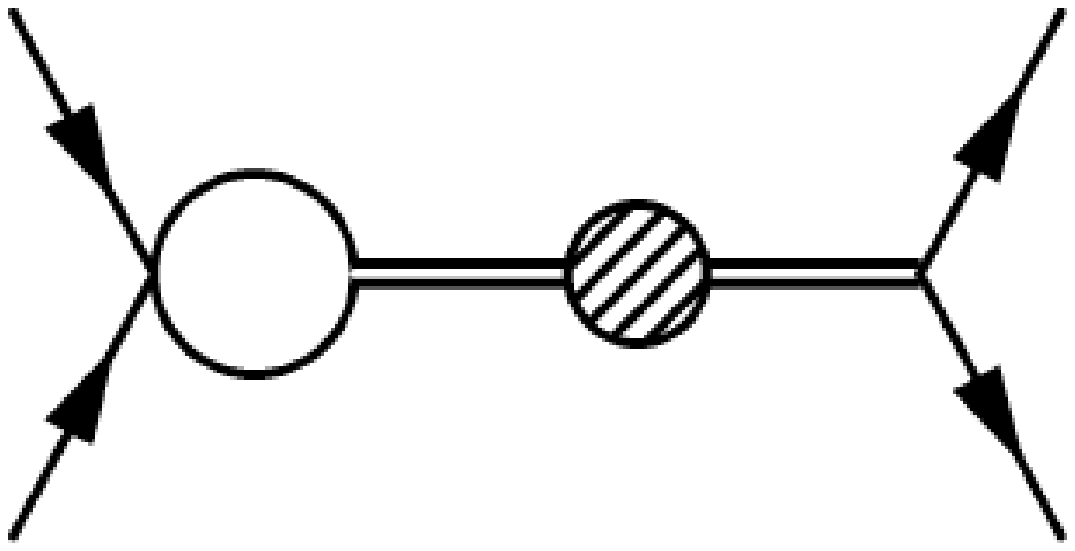}}& \quad
\resizebox{5cm}{2.75cm}{\includegraphics{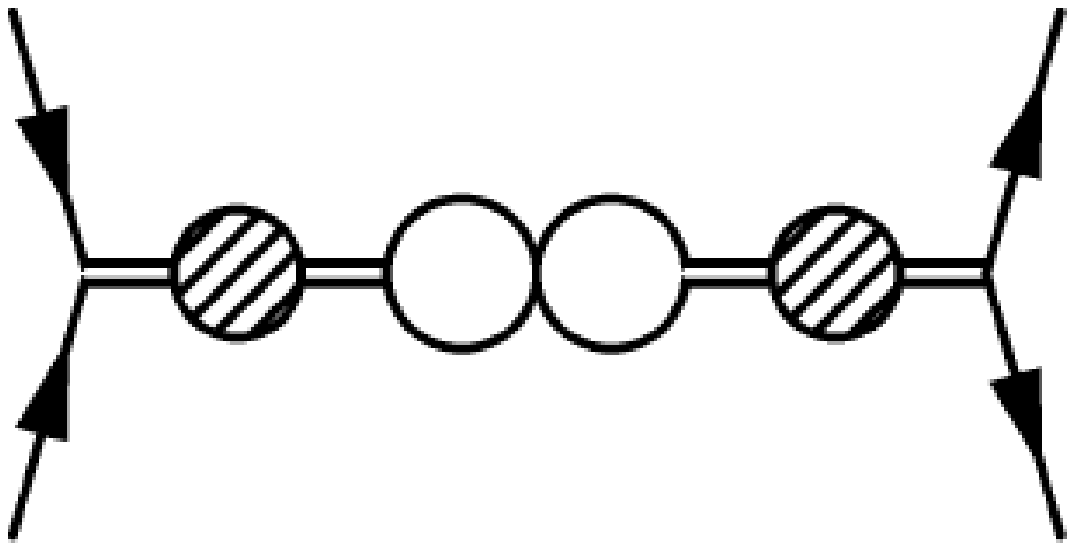}} \\
&\quad(b)&\quad \\ 
\end{tabular}}
\caption{{\footnotesize NLO diagrams. a) Diagrams with a one to one correspondence with the ones in the KSW approach. b) Extra diagrams that 
have been considered in our approach.}}
\label{fig2}
\end{figure}  

At NLO we have the diagrams in Fig.2. The diagrams in Fig.2a have a one to one correspondence with the ones in the KSW approach. We have checked that they give the same contribution as in ref. \cite{Kaplan:1998we}.
The diagrams in Fig.2b are new, and produce the only computational difference between the KSW approach and ours at this order. The insertion of three momentum dependent four nucleon interaction in the bubble chain in KSW corresponds to the tree level energy dependence of the dibaryon field, which becomes NLO in our approach as well, and the insertion of a quark mass dependent four nucleon interaction in the bubble chain corresponds to the dibaryon mass shift due to (\ref{shift}), which may become LO in our approach. Since the diagrams of Fig.2b do not contribute to the $^3S_1- ^3D_1$ mixing, we obtain the same expressions as KSW for it, and hence we will not discuss it further. The contributions of Fig.2b to the scattering amplitudes in the $^1S_0$ and $^3S_1$ channels read
\be
\begin{split}
&\mathcal{A}^{i}_0=C_i+iC_i\Bigl(\frac{m_Np}{2\pi}\Bigr)\mathcal{A}^{i}_{-1}-C_i\Bigl(\frac{m_Np}{4\pi}\Bigr)^2(\mathcal{A}^{i}_{-1})^2 \qquad i=s,v ,
\label{extraamp}
\end{split}
\ee
where $C_s=C_S-3C_T$, $C_v=C_S+C_T$, which produce the following extra contributions to the phase shifts with respect to the KSW approach 
\be
\begin{split}
&\delta_0^{i}=C_i\frac{\frac{m_Np}{4\pi}}{1+\Bigl(\frac{A^2_im_Np}{\pi\d_{m_i}}\Bigr)^2} \qquad i=s,v, 
\end{split}
\label{extraphase}
\ee

\section{Low energy constants from data}
\label{lecs}

In this section we compare the output of our calculation with data and extract the low energy constants. This is important for the self-consistency of the approach: if data favors $\delta_{m_i}$ of the order of the pion mass or smaller, then the introduction of the dibaryon field makes a lot of sense. If, on the contrary, it delivers $\d_{m_i}$ larger than the pion masses, then the introduction of the dibaryon field should be irrelevant. In the first case, if, in addition, $\d_{m_i}\ll m_\pi$ our approach is expected to produce qualitative differences with respect to KSW, at least as far as the quark mass dependence is concerned.

As discussed in the previous section, our final results for the NLO amplitudes and phase shifts may be obtained from those of the KSW approach, by removing the PDS subtractions and adding 
(\ref{extraphase}). It turns out that this extra contribution (\ref{extraamp}) can be written in a form which resembles the quark mass insertion in KSW
(see (\ref{rw}) below). As a consequence,
the following correspondence between our parameters and, for instance, those of ref. \cite{Kaplan:1998we} exist once the PDS subtractions are removed,
\be
\frac{4A_i^2}{\d_{m_i}}=C^i_0 \quad , C_i=-D_2^im_{\pi}^2 \quad ,\frac{4A_i^2}{\d_{m_i}}\frac{1}{m_N \d_{m_i}}=C_2^i,
\ee
Note that whereas there is indeed a one-to-one correspondence between our parameters and the ones of the KSW approach, the quark mass dependence of these parameters does not match, which may become important for eventual extrapolations of lattice data.
 
The fits to phase shift data obtained in this way are very unstable, specially for the $^1S_0$ channel. It is convenient to write these amplitudes in a manifestly unitary form \cite{Oller:2003px,Oller:2000ma}, which is equivalent to the one described above at the order we are working. This is achieved by writing,
\bea
\begin{split}
&\mathcal{A}^{i}_0= \frac{C_i}{\bigl(\frac{4A^2_i}{\d_{m_i}}\bigr)^2} \bigl(\mathcal{A}^{i}_{-1}\bigr)^2 \qquad i=s,v, 
\end{split}\label{rw}
\eea
and proceeding analogously for 
the pion contributions arising from fig.\ref{fig2}a,
\be
\begin{split}
&\mathcal{A}^{i}_{0,\pi}= \frac{\Pi_i}{\bigl(\frac{4A^2_i}{\d_{m_i}}\bigr)^2} \bigl(\mathcal{A}^{i}_{-1}\bigr)^2 \qquad i=s,v, 
\end{split}
\ee
where $\Pi_i$ , $i=s,v $ stand for,

\be
\begin{split}
\Pi_i=\frac{g_A^2}{2f_{\pi}^2}\Biggl\{&\Bigl(\frac{A^2_im_Nm_\pi}{\pi\d_{m_i}}\Bigr)^2\Bigl(1-\frac{1}{4}\ln\Bigl(1+\frac{4p^2}{m_{\pi}^2}\Bigr)\Bigr)-\frac{m_{\pi}^2}{p^2}\Bigl(\frac{A^2_im_Np}{\pi\d_{m_i}}\Bigr)\tan^{-1}\Bigl(\frac{2p}{m_{\pi}}\Bigr)+\\
&+
\frac{m^2_\pi}{4p^2}\ln\Bigl(1+\frac{4p^2}{m_{\pi}^2}\Bigr)-1\Biggr\}.
\end{split}
\label{pifunc}
\ee

Due to the simple selfenergy structure, it is easy to recast these contributions into manifestly unitary expressions, which lead to the following phase shifts\footnote{The unitarization is only approximate for the $^3S_1$ channel, since coupled channel effects are neglected \cite{Oller:2003px,Oller:2000ma}.},

\be
\begin{split}
&\delta_{i}=-\tan^{-1}\Biggl[\frac{  \Bigl(\frac{A^2_im_Np}{\pi\d_{m_i}}\Bigr)}{1-\frac{E}{\d_{m_i}}+\frac{C_i+\Pi_i}{\Bigl(\frac{4A^2_i}{\d_{m_i}}\Bigr)}}\Biggr] \qquad i=s,v.
\end{split}
\label{resum}
\ee
Note that at the order we are working the contribution proportional to $C_i$ can be absorbed into a redefinition of $\d_{m_i}$. This can already be seen at the Lagrangian level. Indeed, a local field redefinition of the dibaryon fields of the type $D \rightarrow D+cNN$, which respects the counting (i.e. $c\sim \Lambda_\chi^{-3/2}$), allows to remove the four nucleon terms at the only cost of introducing higher order operators \cite{Bedaque:1997qi,Bedaque:1999vb,Beane:2000fi}. Nevertheless, we will keep these terms for the fits in this section because the outcome will be illuminating.   
The results discussed in the following as well as the figures correspond to the formula (\ref{resum}) above.

\subsection{Fitting the phase shifts}

Fitting the expressions (\ref{resum}) to the Nijmegen data for the phase shifts instead of the expressions in Section 4 leads to much more stable results, and a very good agreement with data at NLO. We restrict ourselves to energies in the range $0-50 MeV$, for which our lower energy EFT should hold. We present the results of the fit for the LO and the NLO expressions in order to keep track of the evolution of the parameters.
For the $ ^1S_0$ channel at LO we obtain,

\be
{A_s^2\over \d_{m_s}}=-3.20\cdot10^{-5} MeV^{-2},
\ee
and the fit is rather poor. At NLO, however, the fit becomes extremely good (Fig.3), we obtain

\be
\begin{split}
C_s=-1.225\cdot10^{-4} MeV^{-2}\quad ,& \quad A_s=0.0275 MeV^{-1/2}\quad , \quad \d_{m_s}=21.2 MeV, \\
& {A_s^2\over  \d_{m_s}}=3.56\cdot10^{-5} MeV^{-2},
\end{split}
\ee

\begin{figure}
\centerline{\resizebox{12cm}{9cm}{\includegraphics{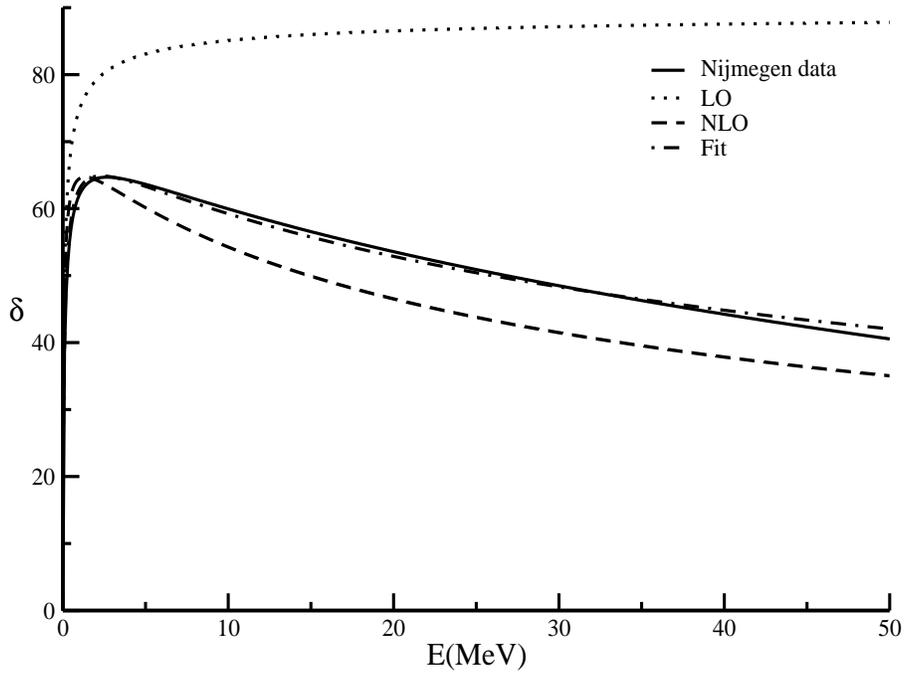}}}
\caption{{\footnotesize The solid line shows the Nijmegen data for the $^1S_0$ phase shift. The doted curve corresponds to fitting at LO the scattering length, while the dashed line corresponds to fitting at NLO the scattering length, the effective range and the first shape parameter ($v_{2s}$). The dot-dashed curve corresponds to fitting $\d$ in the energy range shown.}}
\label{fig3}
\end{figure}

For the $ ^3S_1$ channel at LO we already obtain a reasonable fit with, 
\be
{A_v^2\over \d_{m_v}}=2.57\cdot10^{-4} MeV^{-2},
\ee
which becomes better at NLO (Fig.4) with,

\be\label{fitv}
\begin{split}
C_v=-0.228\cdot10^{-4} MeV^{-2}\quad ,& \quad A_v=0.0332 MeV^{-1/2}\quad , \quad \d_{m_v}=32.0 MeV, \\ 
& {A_v^2\over  \d_{m_v}}=0.344\cdot10^{-4} MeV^{-2},
\end{split}
\ee

We observe that the values for $\d_{m_i}$ are small, namely $\d_{m_i} \sim m_\pi^2/\Lambda_\chi \ll m_\pi$. This explains the large variations of $A_i^2/  \d_{m_i}$ in going from LO to NLO: $\d_{m_i}-\d_{m_i}'$ is not a small correction to $\d_{m_i}'$ in (\ref{shift}) but a quantity of similar size. $A^2_i$ 
take reasonable values $\sim 1/\Lambda_\chi\sim 1/m_N$, and $C_S$ and $C_T$ take values larger than expected with the size assignment $\sim  1/\Lambda_\chi^2$ . In fact, a size assignment $\sim 1/\Lambda_\chi m_\pi$ appears to be more appropriate. Note that such a size assignment would not allow to remove these terms by a local field redefinition which respects the counting, like the one discussed after (\ref{resum}). In fact, if one sets $C_s=0$, the good fit to data for the $^1S_0$ channel is spoiled. We will return to this point in section \ref{discussion}

If we wish to analyze the convergence of the EFT results for the phase shifts, fitting the LECs to data at each order may not be the optimal way to proceed. 
Since the EFT should work the better the lower the energy is, it appears reasonable to us to extract the LECs from the ERE, as advocated by some authors \cite{PavonValderrama:2004nb,PavonValderrama:2003np}.

\subsection{Inputting the ERE parameters}
At LO the scattering lengths $a_s=-23.7fm$ and $a_v=5.42fm$ are sufficient as an input. We obtain
\be
\begin{split}
& {A_s^2\over \d_{m_s}}=-1.61\cdot10^{-3} MeV^{-2}\\
& {A_v^2\over \d_{m_v}}=3.68\cdot10^{-4} MeV^{-2}.
\end{split}
\ee

\begin{figure}
\centerline{\resizebox{12cm}{9cm}{\includegraphics{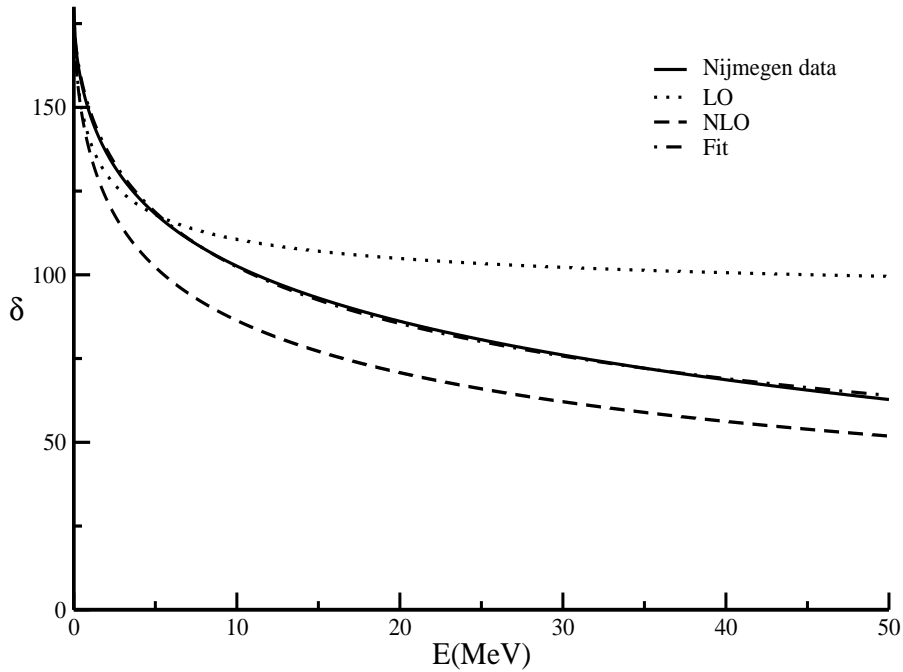}}}
\caption{{\footnotesize The solid line shows the Nijmegen data for the $^3S_1$ phase shift. The doted curve corresponds to fitting at LO the scattering length, while the dashed line corresponds to fitting at NLO the scattering length, the effective range and the first shape parameter ($v_{2t}$). The dot-dashed curve corresponds to fitting $\d$ in the energy range shown.}}
\label{fig4}
\end{figure}

At NLO we need in addition the effective ranges $r_s=2.67fm$ and $r_v=1.83fm$ and the shape parameters $v_{2s}=-0.476fm^3$ and $v_{2v}=-0.131fm^3$\cite{Babenko:2007ss}. We obtain,
\be
\begin{split}
C_s=-0.874\cdot10^{-4} MeV^{-2}\quad ,\quad& A_s=0.0239MeV^{-1/2}\quad , \quad \d_{m_s}=18.9MeV \\
&{A_s^2\over \d_{m_s}}=3.02\cdot10^{-5} MeV^{-2},
\end{split}
\ee
\be
\begin{split}
C_v=-0.768\cdot10^{-4} MeV^{-2}\quad ,\quad& A_v=0.0292 MeV^{-1/2}\quad , \quad \d_{m_v}=19.0 MeV\\
&{A_v^2\over \d_{m_v}}=4.49\cdot10^{-5} MeV^{-2}.
\end{split}
\ee
From the plots in Fig.\ref{fig3} and Fig.\ref{fig4} we see that inputting the parameters from the ERE produces a less satisfactory description of data than the fits. Nevertheless, it might provide more convenient procedure to analyze the convergence of the series.
It also produces smaller values 
of $\d_{m_i}$, which make the contributions of the dibaryon fields more important at low energies.

\section{Rethinking the counting}
\label{discussion}

\begin{figure}
\centerline{\rotatebox{-90}{\resizebox{9cm}{12cm}{\includegraphics{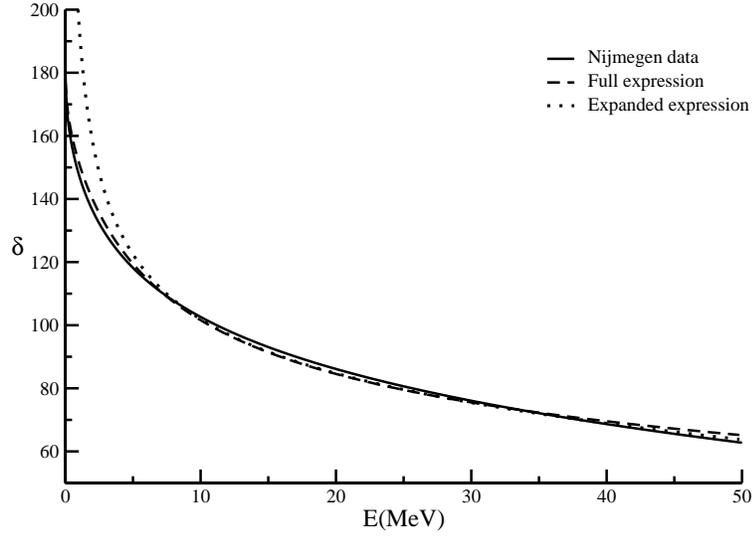}}}}
\centerline{(a)}
\centerline{\rotatebox{-90}{\resizebox{9cm}{12cm}{\includegraphics{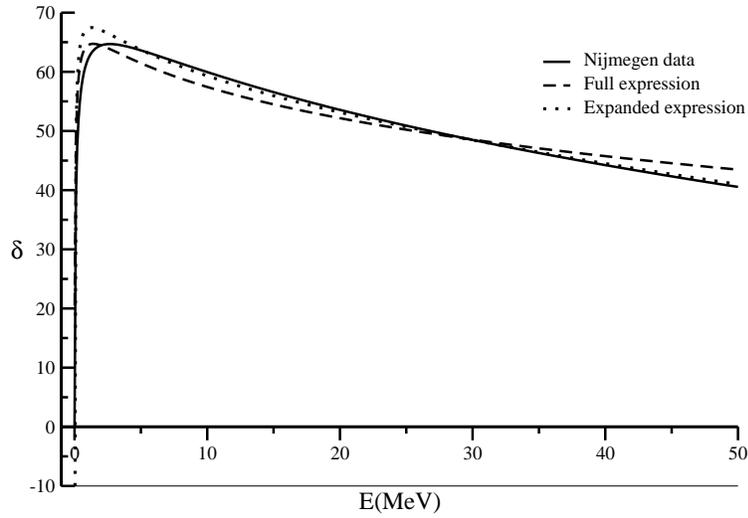}}}}
\centerline{(b)}
\caption{{\footnotesize The solid line shows the Nijmegen data for the phase shifts $^1S_0$ (Fig.a) and $^3S_1$ (Fig.b). The dashed and doted curve correspond to the fit in the range $2-50MeV$ of (\ref{smalldelta}) and its expanded version (as explained in the text) respectively. The expanded expression fails to reproduce the data at very low energies while fits better than the full expression at $p\sim m_{\pi}$. Notice that the expanded expressions for the $^1S_0(^3S_1)$ channel tend to $-\infty(\infty)$ as the energy tends to zero.}}
\label{fig5}
\end{figure}

The numbers for $\d_{m_i}$, $i=s,v$, of tens of $MeV$, clearly indicate that the dibaryon fields must be kept as explicit degrees of freedom essentially at all energies (only for $E\ll \d_{m_i}^2/\Lambda_\chi\sim 1 MeV$ it is justified to integrate them out). In addition, since $\d_{m_i}\sim m_\pi^2/\Lambda_\chi$, it implies that $\d_{m_i}'\lesssim m_\pi^2/\Lambda_\chi$, and hence of size comparable to the NLO contributions. 
This suggests that the actual LO of the dibaryon field propagator should be $\pi/A^2m_Np$ for $p\sim m_\pi$ rather than (\ref{dbself}). This expansion will fail at very low energies, but the manifestly unitary version of it (beyond leading order) is expected to produce sensible results in the very low energy region as well. This can be checked from our expressions (\ref{resum}) by keeping only the dependences in $\d_{m_i}$ which gives rise to LO corrections in the above counting (this is dropping the terms with $C_i$ and keeping only the first term in (\ref{pifunc})), namely, 
\be
\begin{split}
&\delta_{^i}=-\tan^{-1}\Biggl[\frac{  \Bigl(\frac{A^2_im_Np}{\pi\d_{m_i}}\Bigr)}{1-\frac{E}{\d_{m_i}}+\Bigl(\frac{g_A^2A^2_im_N^2m_\pi^2}{8f_{\pi}^2\pi^2\d_{m_i}}\Bigr)\Bigl(1-\frac{1}{4}\ln\Bigl(1+\frac{4p^2}{m_{\pi}^2}\Bigr)\Bigr)}\Biggr] \qquad i=s,v.
\end{split}
\label{smalldelta}
\ee
Fits to data in this case, shown in Fig.\ref{fig5}, are only slightly worse than the ones displayed in Fig.\ref{fig3} and Fig.\ref{fig4}. They deliver,  
\be
\begin{split}
& A_s=0.0305 MeV^{-1/2}\quad , \quad \d_{m_s}=-19.4 MeV \quad , \quad {A_s^2\over \d_{m_s}}=-4.82\cdot10^{-5} MeV^{-2}\\
& A_v=0.0365 MeV^{-1/2}\quad , \quad \d_{m_v}=0.400 MeV \quad , \quad {A_v^2\over \d_{m_v}}=0.00333 MeV^{-2},\\
\end{split}
\label{smalldeltanumbers}
\ee
From the numbers above, 
it is clear that this expansion is expected to work better for the $^3S_1$ channel than for the $^1S_0$ one.
It is also interesting to note that the numbers obtained for $C_i$ from the fits in section \ref{lecs}, namely $C_i\sim 1/\Lambda_\chi m_\pi$, produce
in (\ref{resum})
the expected size for a correction ${\cal O}(m_\pi/\Lambda_\chi )$ to the formula (\ref{smalldelta}) above. Then the $C_i$ in (\ref{resum}) appear to be simulating next order corrections to the phase shifts. This is also so for the second term in (\ref{pifunc})  (the last terms in it give rise to a next-to-next order corrections). For the $^1S_0$ channel this is specially important: if one sets $C_s=0$ in (\ref{resum}) but retains the second term (or the full expression) in (\ref{pifunc}), no good fit to data is achieved. Hence, both terms must be consistently neglected (or taken into account).

In order to evaluate the impact of recasting the amplitude in a manifestly unitary expression, let us expand the formula (\ref{smalldelta})
around $\d_{m_i}=0$ up to NLO terms, \textit{i.e.}, $\tan^{-1}(x)\approx\frac{\pi}{2}-\frac{1}{x}$, and fit this expression to data in the energy range $2-50 MeV$. The expanded expressions still produce very good fits in this energy range (see Fig. \ref{fig5}) and deliver similar values for the parameters,
\be
\begin{split}
& A_s=0.0343 MeV^{-1/2}\quad , \quad \d_{m_s}=-23.8 MeV \quad , \quad {A_s^2\over \d_{m_s}}=-4.96\cdot10^{-5} MeV^{-2}\\
& A_v=0.0367 MeV^{-1/2}\quad , \quad \d_{m_v}=0.404 MeV \quad , \quad {A_v^2\over \d_{m_v}}=0.00333 MeV^{-2},\\
\end{split}
\label{smalldeltanumbers2}
\ee

As mentioned before, this expansion breaks down at very low energies. The manifestly unitary formula (\ref{smalldelta}) essentially fixes up this failure. Alternatively, at these low energies one should change the counting to $p\sim \d_{m_i} \ll m_\pi$. Then one can match our lower energy EFT to an even lower energy EFT with only contact interactions between nucleons (rather than finite range potentials like Yukawa's), and between nucleons and dibaryons. In fact it is within this very low energy EFT that dibaryon fields have mostly been considered in the literature \cite{Bedaque:1997qi,Beane:2000fi,Ando:2004mm}.
It would be interesting to have the explicit relation between the parameters in our fundamental theory and the ones in this very low energy EFT, which is left for future work.

\section{Discussion}

Let us first elaborate on the apparent contradiction between our hypothesis that dibaryon fields must be considered fundamental degrees of freedom and Weinberg's claim that the deuteron is not an elementary particle \cite{wein}.
The point is that one should not identify the $^3S_1$ dibaryon with the deuteron.
We can substantiate this statement by computing the value of $Z=\vert \bra{\Omega} {\vec D}_v \ket{d}\vert^2$ ($d$ stands for the deuteron), the projection of the dibaryon field on the deuteron, from the $LO$ dibaryon propagator:

\begin{equation}
\int d^4x e^{ip\cdot x}\bra{\Omega}T\{D_v^i(0){D_v^j}^\dag(x)\} \ket{\Omega}\vert_{E=p^0\rightarrow E_{d}}=
\frac{iZ\d^{ij}}{E-E_{d}+i\epsilon}+\dots,
\end{equation}
where $E_d$ is the binding energy of the deuteron. At LO
\begin{equation}
\begin{split}
&\int d^4x e^{ip\cdot x}\bra{\Omega}T\{D_v^i(0) {D_v^j}^\dag(x)\} \ket{\Omega}=\frac{i\d^{ij}}{\delta_m+i\frac{A^2m_Np}{\pi}}=\frac{i\d^{ij}}{\delta_m-\frac{A^2m_N}{\pi}\sqrt{-m_N(E+i\epsilon)}},
\end{split}
\end{equation}
from which one easily obtains
\begin{equation}
\begin{split}
& Z=lim_{E \to E_{d}}\frac{\delta_m+\frac{A^2m_N\sqrt{-m_N(E+i\epsilon)}}{\pi}}{\frac{A^4m^3_N}{\pi^2}}=\frac{\delta_m + \vert \delta_m \vert}{\frac{A^4m^3_N}{\pi^2}} \ll 1.
\end{split}
\end{equation}
Hence the projection of the dibaryon field on the physical state, i.e. the deuteron, is parametrically small, and hence there is no contradiction between Weinberg's statement that the deuteron is mainly a nucleon-nucleon bound state, and considering dibaryons as basic degrees of freedom. This is possible because the interaction of the dibaryon with the nucleons is a leading order effect.

Let us next discuss
the quark mass dependence of the scattering lengths $a$ in our approach. At leading non-vanishing order they read 
\be
{m_N\over a}\sim c+c'm_q+c''m_q\ln m_q, 
\ee
($c$, $c'$ and $c''$ stand for quantities which do not depend on the quark masses) in which all three terms are equally important (the chiral log arises from the counterterm of the rightmost diagram in Fig.2a). The most important correction to this formula comes from a particular momentum region, uncovered in \cite{Mehen:1999hz} and recently discussed in \cite{Mondejar:2006yu}, of two loop self-energy diagrams of the dibaryon fields which produces terms proportional to $m_q^{5/4}$ . Hence, the lattice results for the scattering lengths \cite{Fukugita:1994ve,Beane:2006mx} are expected to be very sensitive to the quark masses used. 

Let us finally mention that the inclusion of dibaryon fields as fundamental degrees of freedom may help understanding certain enhancements in $N$-body forces. In the Weinberg's original counting \cite{Weinberg} $N$-body forces were suppressed by powers of $\Lambda_\chi^{4-3N}$. However, the introduction of dibaryon field allows to introduce local terms in the baryon number sector $N$ which are only suppressed by $\Lambda_\chi^{4-3N/2}$ if $N$ is even or $\Lambda_\chi^{(5-3N)/2}$ if $N$ is odd, which indicates that $N$-body forces are expected to be enhanced with respect to Weinberg's counting. In order to illustrate it, consider, for instance, a term 
$r D^\dagger N^\dagger N D$, $r\sim 1/\Lambda_\chi^2$. If the dibaryon field is integrated out and the outcome is expanded considering the momenta small, a six nucleon contact term $r' N^\dagger N^\dagger N^\dagger N N N$, $r' \sim 1/( \Lambda_\chi^3 \d_m^2 )$ is induced, which is indeed enhanced with respect to the Weinberg's counting ($\d_m \ll  \Lambda_\chi$). It is interesting to note that this term is of the same 'unnatural' size as the one found in \cite{Bedaque:1999ve}, if the cut-off $\Lambda$ and the coupling constant $g$ are taken of natural size, namely $\Lambda\sim \Lambda_\chi$ and $g^2\sim 1/\Lambda_\chi$, in that reference. 

\section{Conclusions}

We have proposed a NNEFT at the energy scale of the pion mass, which has two dibaryon fields as explicit degrees of freedom in addition to the nucleons and pions. We have matched it to a lower energy effective theory in which the pion fields have been integrated out. We have pointed out that different counting rules are required for $p \lesssim m_\p^2/\Lambda_\chi$ and for $p\sim m_\p$, and have focused on the latter case. Both effective theories 
are renormalizable and give rise to manifestly unitary amplitudes at leading order. We have calculated these amplitudes at NLO and showed that they produce a very good description of the phase shift in the energy range $5-50 MeV$. Once unitarized the good description extends to the full range $0-50 MeV$. The residual masses delivered by the fits are small, which indicates that the dibaryon fields must also be kept as  explicit degrees of freedom at low energies.


\section*{Acknowledgments}
\indent
We are grateful to Daniel Phillips for his comments on the first version of this paper and for bringing to our attention ref. \cite{Bedaque:1999vb}. We also thank Shung-ichi Ando for bringing to our attention refs. \cite{Ando:2004mm,Ando:2005dk,Ando:2005cz}.
We acknowledge financial support from MEC (Spain) grants CYT FPA 2004-04582-C02-01, FPA2007-60275/, FPA2007-66665-C02-01/ the CIRIT (Catalonia) grant 2005SGR00564, and the RTNs Flavianet MRTN-CT-2006-035482 (EU). JT has been supported by a MEC Collaboration Fellowship and by a FI grant from Ag\`encia de Gesti\'o d'Ajudes Universitaries i de Recerca (AGAUR) of the Generalitat de Catalunya.

\appendix
\renewcommand{\thesection}{\Alph{section}.}
\renewcommand{\theequation}{\thesection \arabic{equation}}

\end{document}